\documentclass[prc,twocolumn,showpacs,amsmath,amssymb]{revtex4}
\usepackage{rotating}
\usepackage{graphicx}
\usepackage{dcolumn}
\usepackage{amstext}
\usepackage{graphicx}
\usepackage{lscape}
\usepackage{epstopdf}
\usepackage{verbatim}
\usepackage[colorlinks,citecolor=blue,bookmarks]{hyperref}
\usepackage{times}
\usepackage{color}
\usepackage{multirow}
\usepackage{hhline}
\usepackage{tabularx}
\usepackage[normalem]{ulem}

\begin{document}

\title{Ternary fission within the temperature dependent relativistic mean field approach}
 
\author{M.T. Senthil kannan$^{1}$}
\email{senthilthulasiram@gmail.com}
\author{Bharat Kumar$^{2,4}$}
\author{M. Balasubramaniam$^{1}$}
\author{B. K. Agrawal$^{3,4}$}
\author{S. K. Patra$^{2,4}$}
\email{patra@iopb.res.in}

\affiliation{$^{1}$Department of Physics, Bharathiar University, Coimbatore - 641046, India.}
\affiliation{$^{2}$Institute of Physics, Sachivalaya Marg, Bhubaneswar - 751005, India.}
\affiliation{$^{3}$Saha Institute of Nuclear Physics, 1/AF,  Bidhannagar, Kolkata - 700064, India.}
\affiliation{$^{4}$Homi Bhabha National Institute, Anushakti Nagar, Mumbai - 400094, India.}
\date{\today}

\begin{abstract}
For the first time, we apply the temperature dependent relativistic
mean field (TRMF) model to study the ternary fission of heavy nucleus
using level density approach. The probability of yields of a particular fragment is obtained by evaluating the convolution integrals which employ the excitation energy
and the level density parameter for a given temperature  calculated
within the  TRMF formalism. To illustrate, we have considered the
ternary fissions in $^{252}$Cf, $^{242}$Pu and $^{236}$U with fixed
third fragment A$_3 = ^{48}$Ca, $^{20}$O and $^{16}$O respectively.
The relative yields are  studied for the temperatures $T =$ 1, 2
and 3 MeV. For the comparison, the relative yields are also
calculated from the single particle energies of the finite range
droplet model (FRDM). In general,  the larger phase space for the
ternary fragmentation is observed indicating that such fragmentations
are most probable ones. For $T =$ 2 and 3 MeV,  the Sn $+$ Ni $+$ Ca
is the most probable combination for the nucleus $^{252}$Cf.  However,
for the nuclei $^{242}$Pu and $^{236}$U, the maximum fragmentation yields
at $T =$ 2 MeV differ from those at $T =$ 3 MeV. For $T =$ 3 MeV, the
closed shell ($Z$ = 8) light mass fragments with
its corresponding partners has larger yield values. But, at  $T =2$ MeV
Si/P/S are  favorable fragments with the corresponding partners.
It is  noticed that the symmetric binary fragmentation along with the
fixed third fragment  for $^{242}$Pu and $^{236}$U are also favored at
$T =$ 1 MeV. The temperature dependence of
the nuclear shape and the single particle energies are  also discussed.

\end{abstract}
\maketitle

\section{Introduction}

The exotic decay modes other than basic decay modes of heavy nuclei
are needed to be studied to understand the reaction kinematics and
the structure as well. One such exotic fission mode of heavy nuclei
is the splitting into three charged fragments so-called ternary fission.
After the earlier reports on ternary fission \cite{pres41,tst47},
the extensive experimental studies on the heavy nuclei $^{252}$Cf,
$^{242}$Pu and $^{236}$U were reported \cite{muga67,ram98,kos99,pya10}. The
observations indicate that alpha particle have the larger yield
values. K\"{o}ster \textit{et. al.} \cite{kos99} reported the ternary
fission yields of $^{242}$Pu for the various third fragment isotopes
up-to $^{30}$Mg. Pyatkov \textit{et. al.} \cite{pya10,pya12} reported
the ternary fission yields of $^{252}$Cf (sf) and $^{236}$U ($n_{th}$,
f) using the missing mass approach. The Sn+Ni/Ge+Ca/S are the most
favorable combinations.  But theoretically, Fong \cite{fon71} calculated
the probability of $\alpha$ particle accompanied fission using statistical
theory. Diehl \textit{et. al.} \cite{die74} applied the liquid drop model
to study of the true ternary fission (TTF) where the three fragments
are almost equal  by direct prolate/oblate and cascade ternary fission
modes. The authors reported that prolate mode is energetically more
favorable than the oblate fission mode. Rubchenya \textit{et. al.}
\cite{rub88} applied the dynamical model for the ternary fission
and reported the formation of light charged particle (LCP) at later
descent stage from the saddle to scission point. Oertzen and Nasirov
\cite{oet14} obtained the TTF fragments using the potential energy
surface (PES) calculations. Manimaran\textit{ et. al.} \cite{mani09}
proposed the three cluster model (TCM) to study the $\alpha$ particle
ternary fission. The obtained relative yield are very well in agreement
with the experimental data. Further, TCM is applied to the study of
equatorial and collinear configuration \cite{vij2014} of all possible
third fragments. The collinear configuration is more favorable for the
heavy third particle accompanied fission with the third fragment at the
middle of two fragments. Rajasekaran and Devanathan \cite{mrd81} applied
the statistical theory to study the binary mass distributions using the
single particle energies of the Nilsson model. The obtained results were
well in agreement with the experimental data. As the sequel of this work,
Balasubramaniam \textit{et. al.} \cite{mbs2014} studied the ternary mass
distribution of $^{252}$Cf for the fixed third fragment $^{48}$Ca using
the single particle energies of the finite range droplet model (FRDM)
and obtained the Sn + Ni + Ca as the most favorable combination at $T =$ 2
MeV. Further, the authors extended \cite{sk16} the study to calculate the
ternary charge distribution of potential energy minimized possible fragments whose probability were calculated using the convolution integrals. The results indicate that
Sn is the one of the most favorable combination for temperature
$T =$ 2 MeV.  The excitation energies and the level density parameters
for different fission fragments required to evaluate the convolution
integrals in such calculations were obtained using temperature independent
single-particle energies from the FRDM corresponding to the ground state
deformations. The temperature dependence was incorporated through the
Fermi-Dirac distributions.

The single-particle energies are usually sensitive to the temperature in
heavy nuclei, in particular, due to the transition from the deformed to
the spherical shape and the transition from pairing phase to the normal
phase as induced by  temperature.  Such features can significantly
affect the temperature dependence of the excitation energy and the
level density parameter. The temperature induced effects on the
nuclear deformation and the pairing phase can be  readily accounted for  within
the temperature dependent non-relativistic and relativistic mean-field
models in self-consistent manner. Of the main focus in the present
investigation is the relativistic mean-field models (RMF). The RMF
models at zero temperature \cite{wal74,seort86,horo 81,gam90,patra91} with
various parameter sets have successfully reproduced the bulk properties,
such as binding energies, root mean square radii, quadrupole deformation
etc. not only for nuclei near the $\beta$ stability line but also for
nuclei away from it.  The temperature dependent relativistic mean field
(TRMF) model has been employed to study the  structural properties of
the highly excited hot nuclei \cite{gam00}. The heavy and rare earth
nuclei are studied within the TRMF model  \cite{bka00,sil01} which indicate
that there is a phase transition from the pairing phase to normal phase
around the temperature $T \sim$ 0.8 MeV and shape transition from prolate
to spherical shape at critical temperature $T_c \sim$ 2.7 MeV.

The RMF formalism is successfully applied to the study
of clusterization of the known cluster emitting heavy nuclei
\cite{aru05,bks06,skp07}. The presence of $\alpha$-clusters in light nuclei,
such as $^{12}$C, which is also an experimental fact is explained very
convincingly within the framework of RMF approximation. In addition, it is
claimed that the N $\neq$ Z clusters exit in the excited states of
heavy nuclei. For superheavy nuclei, the
existence of N $\approx$ Z matter is predicted by this theory. The ternary cluster decay from the hyper-hyper deformed $^{56}$Ni at high angular momenta which is formed in the $^{32}$S $+ ^{24}$Mg reaction is reported in Ref. \cite{beck08}. The RMF model predicted the two multiple $N = Z$, $\alpha$ like clusters or symmetric fission mode of hyper-hyper deformed $^{56}$Ni nucleus \cite{rkg08} which is in contradiction with the experimental results. However, the multiple $\alpha-$nucleus clusterization is in agreement with earlier experiments. Rutz \textit{et. al.} \cite{rutz95} reproduced the double,
triple humped fission barrier of $^{240}$Pu, $^{232}$Th and the asymmetric
ground states of $^{226}$Ra using RMF formalism. Moreover, the symmetric
and asymmetric fission modes are also successfully reproduced. Patra \textit{et. al.} \cite{skp10} studied the neck configuration in the fission decay
of neutron rich U and Th isotopes. Further, various decay modes, such
as $\alpha$-decay, $\beta$-decay and cluster decays are studied in
Refs. \cite{bbs11,bir12,sahu14,kumar15,bhar15} using RMF formalism with
double folding M3Y, LR3Y and NLR3Y nucleon-nucleon interaction potential
within the preformed cluster model.

In the present work we studied the ternary fission of heavy nuclei
$^{252}$Cf, $^{242}$Pu and $^{236}$U  using the temperature dependent
relativistic mean field (TRMF) model.  The various inputs, like, single
particle energies, excitation energies and the level density parameters
of the fission fragments are calculated using the  TRMF model with the
well known NL3 parameter set \cite{lala97}.  For comparison, we calculate
the ternary mass distributions using the single particle energies of
FRDM as explained in Ref. \cite{sk16}.

The article is organized as follows. Section \ref{sec2} provides a brief
description of statistical theory and the TRMF with inclusion of BCS
pairing  formalism used for this study. In section \ref{sec3} we present
our total energy calculations and the temperature dependence of the
excitation energies, $\beta_2$, single particle energies. Further,
we discuss about the ternary mass distribution of heavy nuclei and
the temperature dependence of level density parameter and the level
density. The main results are summarized in Sec.  \ref{se4}.

\section{Formalism} \label{sec2}
We generate different combinations of ternary fission fragments by
considering their charge to mass ratio to be equal to that of the parent
nucleus \cite{mrd81,mbs2014}  i.e.,
 \begin{equation}
\frac{Z_P}{A_P} \approx \frac{Z_i}{A_i} \label{eq1}
\end{equation}
where $A_P$, $Z_P$ and $A_i$, $Z_i$ ($i$ = 1, 2, and 3) correspond to mass
and charge number of the parent nucleus and three fission fragments, respectively. The
following constraints, $A_1 + A_2 + A_3= A$, $Z_1 + Z_2 + Z_3 = Z$,
and $ A_1 \ge A_2 \ge A_3 $ are imposed to satisfy the conservation of
mass and charge number in nuclear reaction and to avoid the repetition
of fragment combinations. The third fragment $A_3$ is also considered
a priori to find the other two fragments $A_1$ and $A_2$.

\subsection{Statistical Theory} \label{ST}
According to the statistical theory\cite{fon56,mor72,mrd81,sk16}, the
ternary fission probability $P(A_j,Z_j)$ is proportional
to the folded densities $\rho_{123}(A_i,Z_i,E^*)$ of the three distinct
fragments and is given by,

\begin{widetext}
\begin{equation}
\rho_{123}(A_i,Z_i,E^*) = \int_{0}^{E^*} \rho_{1}(A_1,Z_1,E^*_1)\left[\int_{0}^{E^*} \int_{0}^{E^*}
\prod_{i=2}^{3} \rho_i(A_i,Z_i,E_i^*)
\delta(E_2^*+E_3^*-(E^*-E_1^*))dE^*_i \right]dE^*_{1}, \label{eq2}
\end{equation} 
\end{widetext}
with $E^{*}$ as the excitation energy. Here, $\rho_i$ is the level density
of three fragments ($i$ = 1, 2, 3). The double integral in the square
bracket is the binary convolution integral. The nuclear level density
\cite{bet37,mor72} is expressed as a function of fragment excitation
energy $E^{\ast}_i$ and the single particle level density parameter
$a_i$ is,

\begin{equation}
\rho_i\left(E^{\ast}_i\right) = \dfrac{1}{12} \left(\dfrac{\pi^{2}}{a_i}\right)^{1/4} E^{\ast (-5/4)}_i\exp\left(2\sqrt{a_iE^{\ast}_i}\right).\label{eq3}
\end{equation}
In the Refs. \cite{mbs2014,sk16}, we calculated the excitation energies
of the fragments using the single particle energies of 
FRDM \cite{moller97} at a given temperature $T$. In
the present work we applied the self consistent temperature dependent
relativistic mean field theory to calculate the excitation energy of
the fragments. The excitation energy is calculated  as,
\begin{equation}
 E^{*}_{i}(T) = E(T) - E(T = 0). \label{eq4}
\end{equation}
The level density parameter $a_i$ is given as,
\begin{equation}
a_i = \frac{E^*_i}{T^{2}}. \label{eq5}
\end{equation}
The relative yield is calculated as the ratio between the probability
of a given ternary fragmentation and the sum of the probabilities of
all the possible ternary fragmentations and it is given by,
\begin{equation}
Y(A_j,Z_j)=\frac{P(A_j,Z_j)}{\sum_{j} P(A_j,Z_j)}. \label{eq6}
\end{equation}
The competing basic decay modes such as neutron emission, $\alpha$
decay, binary fragmentation are not considered in the present work. The
presented results are the prompt disintegration of a parent nucleus
into three fragments (democratic breakup). The resulting excitation energy would
be liberated as prompt particle emission or delayed emission, but such
secondary emissions are not considered in the present study.

\subsection{RMF Formalism}
The RMF theories assume that the nucleons interact with each other
via the meson fields. The nucleon - meson interaction is given by the
Lagrangian density \cite{patra91,wal74,seort86,horo 81, bogu77,pric87},

\begin{eqnarray}
{\cal L}&=&\overline{\psi_{i}}\{i\gamma^{\mu}
\partial_{\mu}-M\}\psi_{i}
+{\frac12}\partial^{\mu}\sigma\partial_{\mu}\sigma
-{\frac12}m_{\sigma}^{2}\sigma^{2}\nonumber\\
&& -{\frac13}g_{2}\sigma^{3} -{\frac14}g_{3}\sigma^{4}
-g_{\sigma}\overline{\psi_{i}}\psi_{i}\sigma\nonumber\\
&&-{\frac14}\Omega^{\mu\nu}
\Omega_{\mu\nu}+{\frac12}m_{w}^{2}V^{\mu}V_{\mu}
 -g_{w}\overline\psi_{i}
\gamma^{\mu}\psi_{i}
V_{\mu}\nonumber\\
&&-{\frac14}\vec{B}^{\mu\nu}.\vec{B}_{\mu\nu}+{\frac12}m_{\rho}^{2}{\vec
R^{\mu}} .{\vec{R}_{\mu}}
-g_{\rho}\overline\psi_{i}\gamma^{\mu}\vec{\tau}\psi_{i}.\vec
{R^{\mu}}\nonumber\\
&&-{\frac14}F^{\mu\nu}F_{\mu\nu}-e\overline\psi_{i}
\gamma^{\mu}\frac{\left(1-\tau_{3i}\right)}{2}\psi_{i}A_{\mu}.
\label{eq:7}
\end{eqnarray}
Where, $\psi_i$ is the single particle Dirac spinor. The arrows over the letters in the above equation represent the isovector quantities. The nucleon, the $\sigma$, $\omega$, and $\rho$ meson masses are denoted by M, $m_\sigma$, $m_\omega$ and $m_\rho$ respectively.
 The meson and the photon fields are denoted by 
$\sigma$, $V_\mu$, $R^\mu$ and $A_\mu$ for $\sigma$, $\omega$, $\rho-$ mesons and photon respectively. The $g_{\sigma}$, $g_\omega$, $g_\rho$ and $\frac{e^2}{4\pi}$ are the coupling constants for the $\sigma$, $\omega$, $\rho-$ mesons and photon fields with nucleons respectively. The strength of the constants $g_2$ and $g_3$ is responsible for the nonlinear coupling of $\sigma$  meson ($\sigma^3$ and $\sigma^4$). The field tensors of the isovector mesons and the photon are given by,
\begin{eqnarray}
	\Omega^{\mu\nu} & = & \partial^{\mu} V^{\nu} - \partial^{\nu} V^{\mu}, \, \\[3mm]
	\vec{B}^{\mu\nu} & = & \partial^{\mu} \vec{R}^{\nu} - \partial^{\nu} \vec{R}^{\mu} - g_{\rho} (\vec{R}^{\mu}\times\vec{R}^{\nu}), \,  \\[3mm]
	F^{\mu\nu} & = & \partial^{\mu} A^{\nu} - \partial^{\nu} A^{\mu}.	\, 
\end{eqnarray}
The classical variational principle gives the Euler-Lagrange equation, we get the Dirac-equation with potential terms for the nucleons and Klein-Gordan equations with source terms for the mesons. We applied the no-sea approximation, so we neglected the antiparticle states. We are dealing with the static nucleus, so the time reversal symmetry and the conservation of parity simplifies the equations. After simplifications, the Dirac equation for the nucleon is given by,
\begin{equation}
	\{ - i\alpha.\bigtriangledown + V(r) + \beta\left[M + S(r)\right]\}\,\,\psi_i = \epsilon_i \, \psi_i, \label{eq11}
\end{equation}
where V(r) represents the vector potential and S(r) is the scalar potential,
\begin{eqnarray}
V(r) &=& g_\omega \omega_0 + g_\rho \tau_{3} \rho_0(r)+ e \frac{(1-\tau_3)}{2} A_0(r) \nonumber \\ [2mm]
S(r) & =& g_\sigma \sigma(r),
\end{eqnarray}
which contributes to  the effective mass,
\begin{equation}
	M^*(r) = M + S(r).
\end{equation}

The Klein-Gordon equations for the meson and the electromagnetic fields with the nucleon densities as sources are,
\begin{center}
\begin{eqnarray}
\{-\triangle + m_\sigma^2\}\sigma(r) = -g_{\sigma}\rho_s(r)  -g_2\sigma^2(r)&-\, g_3\sigma^3(r), \\[2mm] \label{eq14}
\{-\triangle + m_\omega^2\}\omega_0(r) = g_\omega\rho_v(r), & \\[2mm] \label{eq15}
\{-\triangle + m_\rho^2\}\rho_0(r) = g_\rho\rho_3(r), &\\[2mm] \label{eq16}
-\triangle A_0(r) = e\rho_c(3). & \label{eq17}
\end{eqnarray}	
\end{center}
The corresponding densities such as scalar, baryon (vector), isovector and proton (charge) are given as
\begin{eqnarray}
\rho_s(r) & = &
\sum_i n_i \, \psi_i^\dagger(r) \, \psi_i(r) \,,
\label{eqFN6} \\[1mm]
\rho_v(r) & = &
\sum_i n_i \,\psi_i^\dagger(r) \, \gamma_0 \,\psi_i(r) \,,
\label{eqFN7} \\[1mm]
\rho_3 (r) & = &
\sum_i n_i \, \psi_i^\dagger(r)\, \tau_3\, \psi_i(r) \,,
\label{eqFN8} \\[1mm]
\rho_{\rm p}(r) & = &
\sum_i n_i \,\psi_i^\dagger(r) \,\left (\frac{1 -\tau_3}{2}
\right) \, \psi_i(r) \,.
\label{eqFN9} 
\end{eqnarray}
To solve the Dirac and Klein-Gordan equations, we expand
the Boson fields and the Dirac spinor in an axially deformed
symmetric harmonic oscillator basis with $\beta_0$ as the initial deformation
parameter. The nucleon equation along with different meson equations
form a set of coupled equations, which can be solved by iterative
method. The center of mass correction is
calculated with the non-relativistic approximation $E_{c.m.} = -3/4 \times 41A^{-1/3}$. The quadrupole deformation parameter
$\beta_2$ is calculated from the resulting quadrupole moments of the
proton and neutron. The total energy is given by \cite{blunden87,reinhard89,gam90},

\begin{eqnarray}
E(T) =& \sum_i \epsilon_i n_i + E_\sigma + E_{\sigma NL} + E_\omega + E_\rho\nonumber
\\[2mm]& + E_C +E_{pair} + E_{c.m.} - AM,
\end{eqnarray}
with
\begin{equation}
E_\sigma  = -\frac{1}{2}g_\sigma \int d^3r \rho_s(r) \sigma(r),
\end{equation}
\vspace{-3mm}
\begin{equation}
E_{\sigma NL} = -\frac{1}{2}g_\sigma \int d^3r \left\lbrace\, \frac{1}{3}g_2 \,\sigma^3(r) + \frac{1}{2}g_3\, \sigma^4(r)\,\right\rbrace,
\end{equation}
\vspace{-3mm}
\begin{eqnarray}
E_\omega = &  -\frac{1}{2}g_\omega \int d^3r \rho_v(r)\omega^0(r),\\[3mm]
E_\rho= &  -\frac{1}{2}g_\sigma \int d^3r\rho_3(r)\rho^0(r), \\[3mm]
E_C = &  -\dfrac{e^2}{8\pi}\int d^3r\rho_c(r)A^0(r),
\end{eqnarray}
\vspace{-3mm}
\begin{eqnarray}
E_{pair} = - \triangle\sum_{i>0}u_{i}v_{i} = -\frac{\triangle^2}{G}, 
\end{eqnarray}
\vspace{-3mm}
\begin{equation}
E_{c.m.}= -\frac{3}{4}\times 41A^{-1/3}.
\end{equation}
Here, $\epsilon_i$ is the single particle energy, $n_i$ is the occupation probability and $E_{pair}$ is the pairing energy obtained from the simple BCS formalism.

\subsection{Pairing and temperature dependent RMF formalism}\label{sec:bcs}
Pairing correlation plays a pivotal role in the description of the open shell nuclei and the quantitative description of deformation in heavy nuclei. In the Hartree approximation, we have only, ${\psi}^{\dagger}{\psi}$ (density) term in the Lagrangian. The inclusion of pairing term like ${\psi}^{\dagger}{\psi}^{\dagger}$, $\psi\psi$ and two body interaction term ${\psi}^{\dagger}{\psi}^{\dagger}\psi\psi$ violates the particle number conservation. So, we applied externally the BCS constant pairing gap approximation for our calculation to take the pairing correlation into account. The  pairing interaction energy in terms of occupation probabilities $v_i^2$ and $u_i^2=1-v_i^2$ is written 
as~\cite{pres82,patra93}:
\begin{equation}
E_{pair}=-G\left[\sum_{i>0}u_{i}v_{i}\right]^2,
\end{equation}
with $G$ is the pairing force constant. 
The variational approach with respect to the occupation number $v_i^2$ gives the BCS equation 
\cite{pres82}:
\begin{equation}
2\epsilon_iu_iv_i-\triangle(u_i^2-v_i^2)=0,
\label{eqn:bcs}
\end{equation}
with the pairing gap $\triangle=G\sum_{i>0}u_{i}v_{i}$. The pairing gap ($\triangle$) of proton and neutron is taken from the empirical formula \cite{va73,gam90}:
\begin{equation}
\triangle = 12 \times A^{-1/2}.
\end{equation}
 The temperature introduced in the partial occupancies in the BCS approximation is given by,
\begin{equation}
n_i=v_i^2=\frac{1}{2}\left[1-\frac{\epsilon_i-\lambda}{\tilde{\epsilon_i}}[1-2 f(\tilde{\epsilon_i},T)]\right],
\end{equation}
with 
\begin{eqnarray}
f(\tilde{\epsilon_i},T) = \frac{1}{(1+exp[{\tilde{\epsilon_i}/T}])} & and \nonumber \\[3mm]
\tilde{\epsilon_i} = \sqrt{(\epsilon_i-\lambda)^2+\triangle^2}.&
\end{eqnarray}

The function $f(\tilde{\epsilon_i},T)$ represents the Fermi Dirac
distribution function for quasi particle energies $\tilde{\epsilon_i}$. The chemical potential $\lambda_p (\lambda_n)$
for protons (neutrons) is obtained from the constraints of particle number
equations

\begin{eqnarray}
\sum_i n_i^{Z}  = Z, \nonumber \\
\sum_i n_ i^{N} =  N.
\end{eqnarray}
The sum is taken over all proton and neutron states. The entropy is obtained by,
\begin{equation}
S = - \sum_i \left[n_i\, lnn_i + (1 - n_i)\, ln (1- n_i)\right].
\end{equation}
The temperature dependent RMF total energies and the gap parameter are obtained by minimizing the free energy,
\begin{equation}
F = E - TS.
\end{equation}
 In constant pairing gap calculations, for a particular value of pairing gap $\triangle$
 and force constant $G$, the pairing energy $E_{pair}$ diverges, if
 it is extended to an infinite configuration space. In fact, in all
 realistic calculations with finite range forces, $\triangle$ is not
 constant, but decreases with large angular momenta states above the Fermi
 surface. Therefore, a pairing window in all the equations are extended
 up-to the level $|\epsilon_i-\lambda|\leq 2(41A^{-1/3})$ as a function
 of the single particle energy. The factor 2 has been determined so as
 to reproduce the pairing correlation energy for neutrons in $^{118}$Sn
 using Gogny force \cite{gam90,patra93,dech80}.

\section{Results and discussions} \label{sec3}
In earlier studies \cite{mbs2014,sk16}, the level
densities of the fragments were calculated using the single particle energies from
the Finite Range Droplet Model (FRDM) of M\"{o}ller \textit{et. al.}
\cite{moller95}. The single particle levels were retrieved from the
Reference Input Parameter Library (RIPL-3) \cite{ripl3}. In the present
study, we calculate the level densities using the TRMF formalism. Before
embarking on our main results, we discuss about the temperature
induced structural changes in the $^{252}$Cf, $^{242}$Pu and $^{236}$U
nuclei. Next, we calculate relative yields for the ternary fission of
$^{252}$Cf, $^{242}$Pu  and $^{236}$U with the fixed third fragments
A$_3 =\, ^{48}$Ca, $^{20}$O and $^{16}$O respectively. The other two fragments with masses and charges $A_1$, Z$_1$ and A$_2$, Z$_2$ are obtained by keeping the mass to charge ratio to be equal to that of parent nucleus as given by  Eq. \eqref{eq1}.
The results are presented for the three different temperatures $T =$
1, 2 and 3 MeV.  In principle, one should consider all the possible
third fragment. However, in the present study we have neglected such
possibilities.  From the cluster decay study of $^{252}$Cf \cite{mbs99},
it is shown that $^{48}$Ca or the neighboring $^{48}$Ar or $^{52}$Ca
have large preformation probability compared to their light clusters,
such as C, O etc.  In the view of experimental data \cite{kos99} $^{20}$O
is chosen for $^{242}$Pu as the third fragment.

The TRMF equations for the nucleon and the Boson fields are solved
within the basis expansion method. In the present work, the number of
oscillator shells $N_F=12$ and $N_B=20$ are used as the basis space  for
the nucleons and boson fields, respectively. The total energy is obtained
by minimizing the free energy at a given temperature. The ground state
($T =$ 0 ) binding energies are well reproduced with the experimental
data in our calculations.

\subsection{Excitation energies, quadrupole deformation parameter and single particle energies}

\begin{figure}[t]
\includegraphics[width=1\columnwidth]{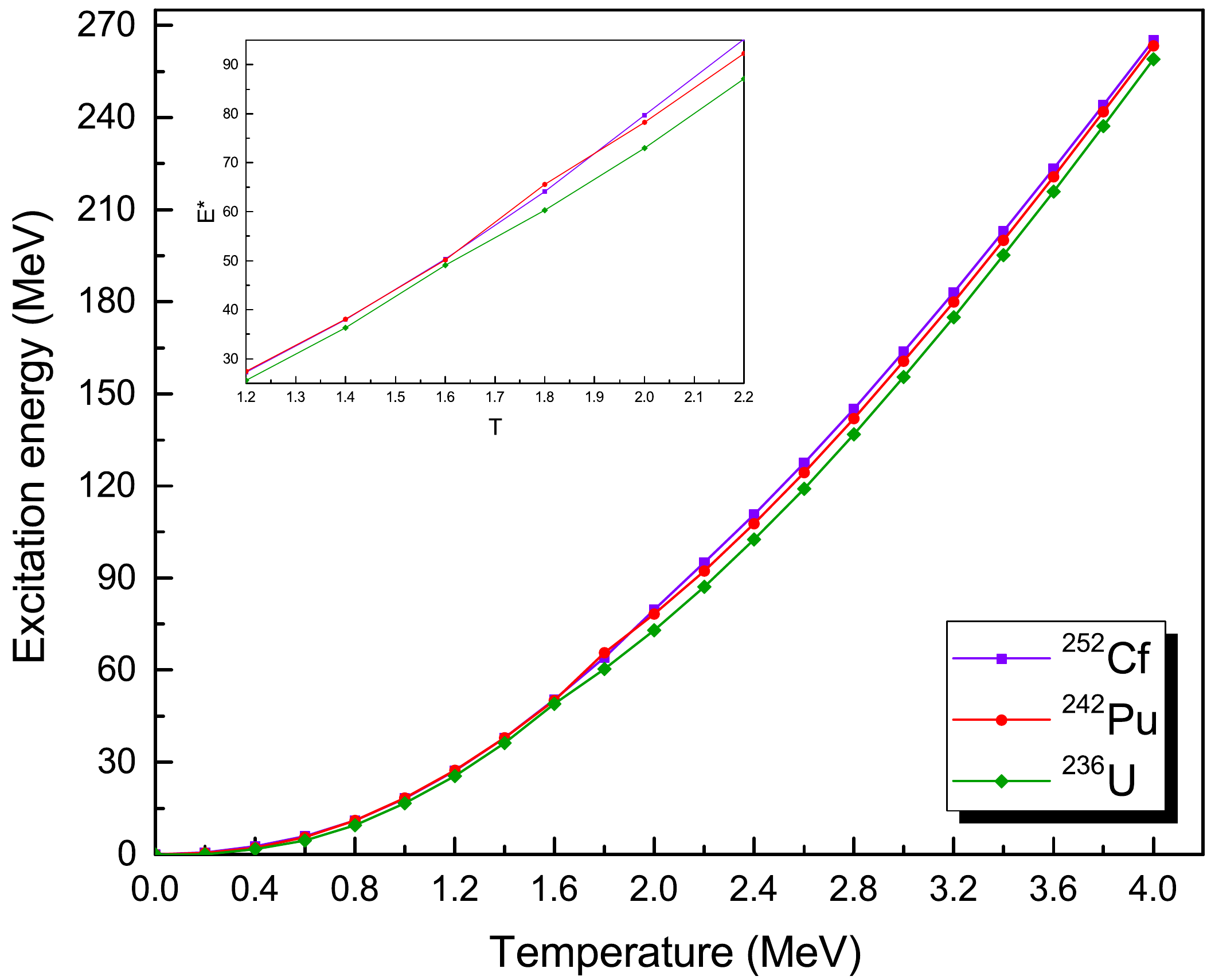}
\caption{(Color online) Temperature dependence of the excitation energies for
the nuclei $^{252}$Cf, $^{242}$Pu and $^{236}$U. 
}\label{excomp} \end{figure}

The shape transitions in  $^{166}$Er and $^{170}$Er using the TRMF
formalism is studied in \cite{bka00}. The shape transition occurs at
$T =$ 2.7 MeV. Similar studies have been performed by applying the finite temperature Hartree Fock Bogoliubov method to the finite range density dependent Gogny force
\cite{egi00} and the pairing  plus quadrupole force \cite{goo86}. These
results are similar to those obtained within the TRMF formalism. At finite
temperatures, the continuum corrections due to the excitation of nucleons
in the continuum are to be considered. The level density in the continuum
depends on the basis space parameter N$_F$ and N$_B$ \cite{niu09}. It
has been shown that the continuum corrections are not very important
in the calculations of level densities up-to the temperature T $\sim$
3 MeV \cite{bka00,bka98}. In our present study, we neglected the continuum
corrections because the considered temperatures are up to T $=$ 3 MeV.
Further, we do not include the thermal fluctuations due to computational
limitations.  The thermal fluctuations are to  be included for the more
quantitative study of shape transitions.

\begin{figure}[b]
\includegraphics[width=1\columnwidth]{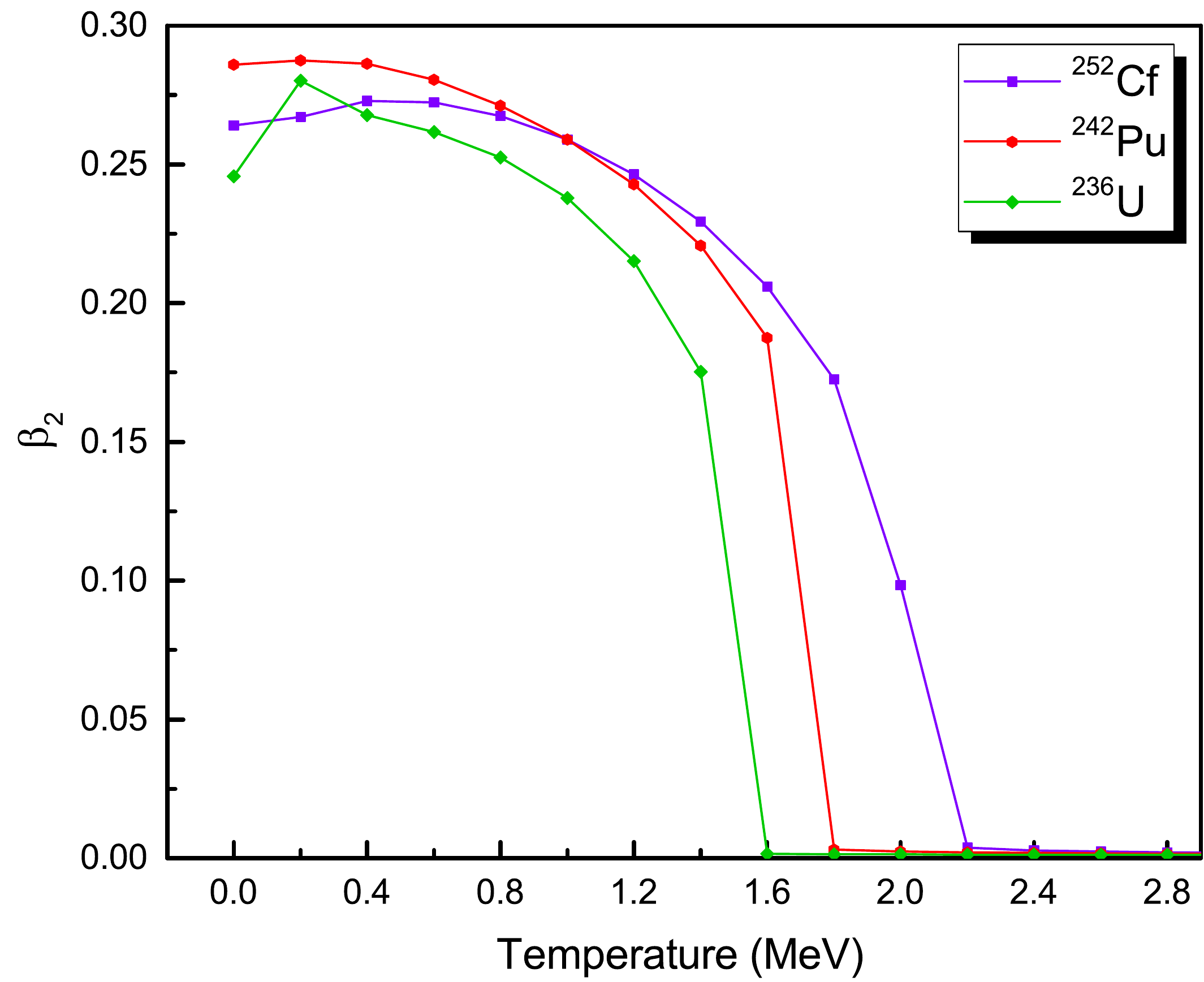}
\caption{(Color online) Temperature dependence of the quadrupole
deformation parameter $\beta_2$ for the nuclei $^{252}$Cf, $^{242}$Pu
and $^{236}$U.} \label{b2comp} 
 \end{figure} 

 \begin{figure*}[!]
\includegraphics[width=1.6\columnwidth]{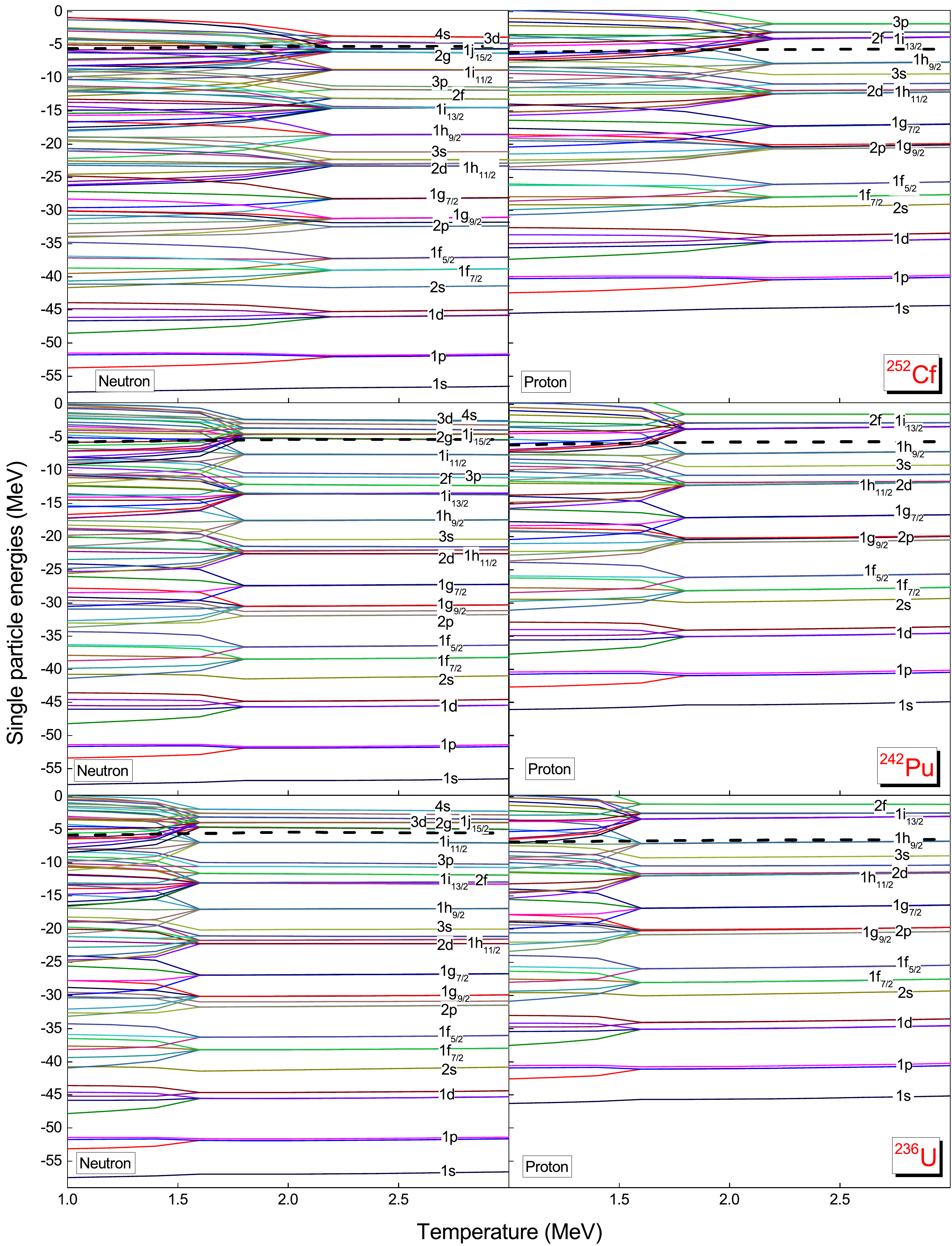} \caption{(Color
online) Variation of single particle levels of $^{252}$Cf, $^{242}$Pu
and $^{236}$U with temperature T. Fermi levels are denoted by the dashed 
line(black).}  \label{spe}
 \end{figure*}

The temperature dependence of the excitation energies of the parent nuclei
$^{252}$Cf, $^{242}$Pu and $^{236}$U are shown in Fig. \ref{excomp} for $T
=$ 0-4 MeV. The excitation energy of the nuclei increases quadratically
with the temperature, as given by Fermi gas approximation $E^{\ast}
= a T^2$. Further, small deviations from the quadratic behavior are
observed for the nuclei $^{242}$Pu and $^{236}$U curves (depicted
inside the Fig. \ref{excomp}) at the temperatures $T =$ 1.8 and 1.6 MeV
respectively. This is due to the shape transition of the nuclei at these
temperatures which is called critical temperature $T_c$. But, there is no
such deviations seen in the case of $^{252}$Cf. To clarify this we have
plotted the quadrupole deformation parameter($\beta_2$) as a function of
temperature in Fig. \ref{b2comp}. The shape transitions from prolate to
spherical shape occur sharply in $^{242}$Pu and $^{236}$U at the critical
temperatures $T_c$ $=$ 1.8 and 1.6 MeV respectively. In $^{252}$Cf
nucleus the deformation decreases gradually with temperature and it
vanishes at $T_c =$ 2.2 MeV.  For the nuclei $^{242}$Pu and $^{236}$U,
the $\beta_2$ increases up to $T =$ 0.2 MeV due to the transitions from
the pairing  to normal phase.  The pairing transitions occurs at $T =$
0.4 MeV for the nucleus $^{252}$Cf.  In earlier studies \cite{bka00},
the deformation parameter $\beta_2$ dropped rapidly around the transition
temperature and it leveled off within the interval of  $T \simeq 0.2-0.3
$MeV around the transition temperature.

\par In Fig. \ref{spe} we plot the thermal evolution of the single
particle energies (spe) for the protons and neutrons for the considered
nuclei. The black dashed lines in all panels represent the Fermi
surface. It can be seen that  various Nilsson single-particle energies
become degenerate beyond $T_c$. When temperature increases, more levels
above the Fermi surface are occupied. As a result the occupancies of non
degenerate states evolves in a self consistent manner which drives the
single particle potential towards spherically symmetric one. In other
words, the shell structures vanishes at high temperatures and the nucleus
becomes a perfect liquid drop with degenerate Fermions.

\subsection{Ternary fission mass distribution and the level densities}
Pyatkov and Oertzen group \cite{pya10,pya12,oer08,oert08} experimentally
observed the heavy third fragments from  the new decay mode called
collinear cluster tripartition (CCT), in which the ternary fragments
are collinearly emitted due to the lower Coulomb interaction for this
configuration and at least one of the fragment has the composition with
magic number of nucleon.  Further, Pyatkov \textit{et. al.} \cite{pya10}
reported that the CCT decay of $^{252}$Cf with the ternary $^{48}$Ca
yields of $4.7 \pm 0.2 \times 10^{-3} /$ binary fission and the CCT
decay of $^{236}$U with the ternary $^{34}$Si yields of $5.1 \pm 0.4
\times 10^{-3} /$ binary fission. It was further reported that, this
yield is due to the whole {\it Ni-bump} consisting of some hundreds
of different mass partitions. In Ref. \cite{pya12}, it is  mentioned
that the total yield of $^{68,72}$Ni ions do not exceed $10^{-4} /$
binary fission. However, the yield of each separate ternary partition,
for instance $^{128}$Sn +$^{72}$Ni +$^{52}$Ca can be estimated to be
of the order of $3 \times 10^{-6} /$ binary fission.  It is reported
that the heavy third particle cluster like $^{48}$Ca, $^{50}$Ca has
larger yield values in collinear configuration than the light third
fragment $^{4}$He, $^{10}$Be \cite{mani11}.  Recently \cite{mbs2014}
studied ternary fission mass distribution of $^{252}$Cf using FRDM,
for the fixed third fragment $^{48}$Ca, at the temperatures $T =$ 1 and
2 MeV, revealed that Sn + Ni + Ca as the most favorable combination at
$T =$ 2 MeV. For our investigation, we consider one of the nucleus to be
$^{252}$Cf for the study of ternary fission at the temperatures $T =$ 1, 2
and 3 MeV. The ternary mass distribution of $^{242}$Pu is studied using
the third fragment as  $^{20}$O as suggested by K\"{o}ster \cite{kos99}.
We also studied the ternary fission of $^{236}$U for the fixed cluster
like third fragment $^{16}$O. For the comparison, the ternary mass
distributions are also calculated using the FRDM formalism.

\par The total energy at finite temperature and ground state energy
are calculated using the TRMF formalism as discussed in the section
\ref{ST}. From the TRMF the excitation energy E$^{\ast}$ of fragments are
calculated using Eq. \eqref{eq4}. From the excitation energy E$^{\ast}$
and the temperature $T$ the level density parameter $a$ is calculated
using Eq. \eqref{eq5}. From the excitation energy E$^{\ast}$ and the
level density parameter $a$, the level density $\rho$ of fragments are
calculated using Eq. \eqref{eq3}. From the fragment level densities
$\rho_i$, the folding density $\rho_{123}$ is calculated using the
convolution integral Eq. \eqref{eq2} and the relative yield values are
calculated using Eq. \eqref{eq6}. It is to be noted that, the total
yield values are normalized to 2 throughout the calculations. In FRDM
formalism, the temperature dependence introduced in the Fermi occupation
number. Using the Lagrange multipliers $\alpha^{N,Z}$ and $\beta$ and the
number equations, the temperature dependent energy $E(T)$ is calculated
from the ground state single particle energies for a given temperature
$T$. The excitation energy $E^*$ at the given temperature is $E^* = E(T)
- E(0)$ and other details can be found in Ref. \cite{mbs2014}.

\begin{figure}[t]
	\includegraphics[width=1.\columnwidth]{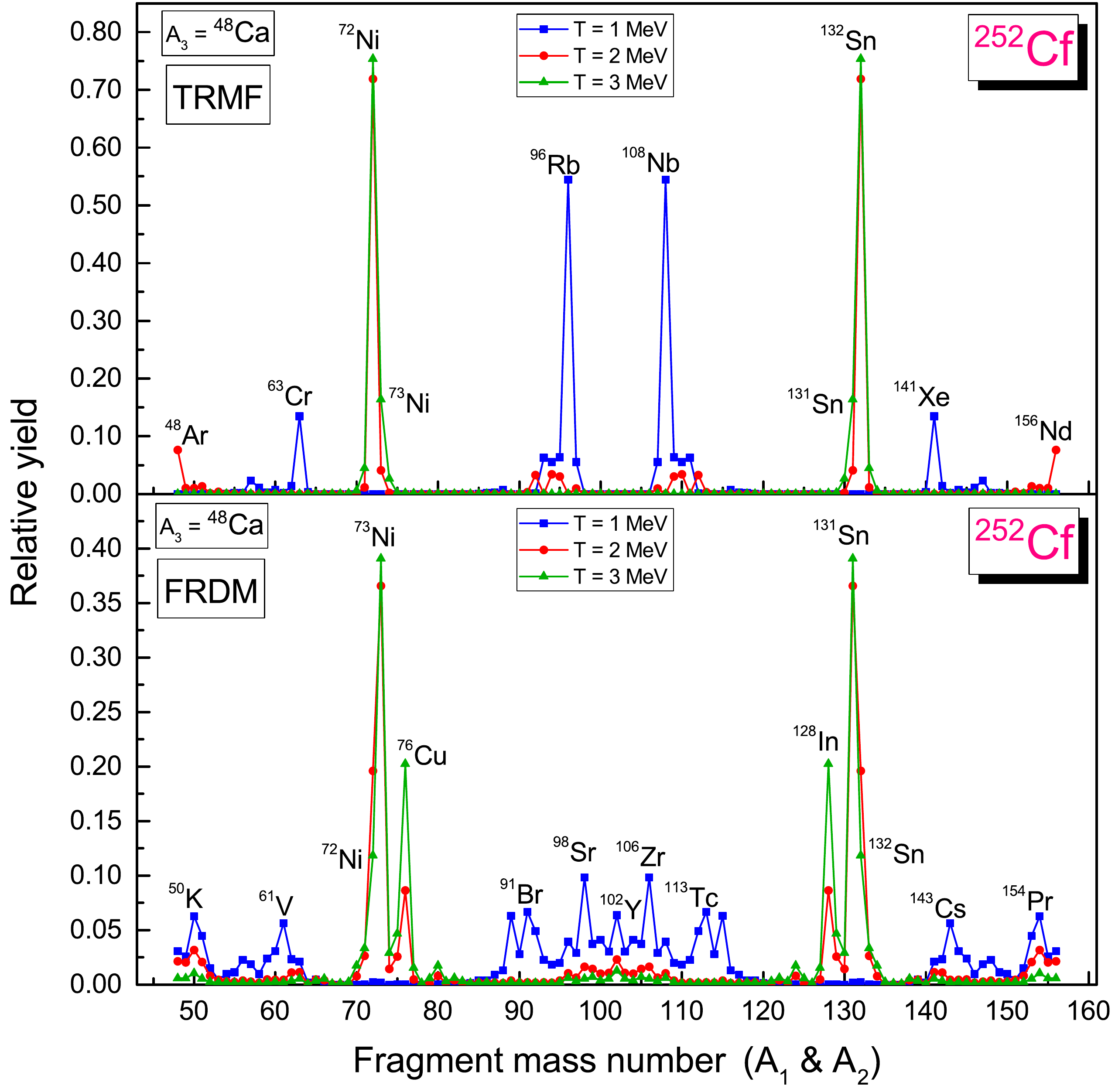}
	\caption{(Color online) Mass distribution of $^{252}$Cf for the fixed third fragment $^{48}$Ca for the temperatures T = 1, 2 and 3 MeV. The total yield values are normalized to 2.}
	\label{Cf_yield}
\end{figure}

In Fig. \ref{Cf_yield}, the TRMF results for the  ternary fission mass distributions of
$^{252}$Cf for the fixed third fragment $^{48}$Ca are shown for different temperatures. For, $T =$ 1
MeV, $^{108}$Nb $+^{96}$Rb $+^{48}$Ca is the most probable fragmentation
followed by emission of $^{141}$Xe $+^{63}$Cr $+^{48}$Ca. For higher
temperatures $T =$ 2 and 3 MeV, it is interesting to see that $^{132}$Sn
$+^{72}$Ni $+^{48}$Ca is the most favorable combination of the existing
fragmentations.
  \begin{figure}[!]
\includegraphics[width=1.\columnwidth]{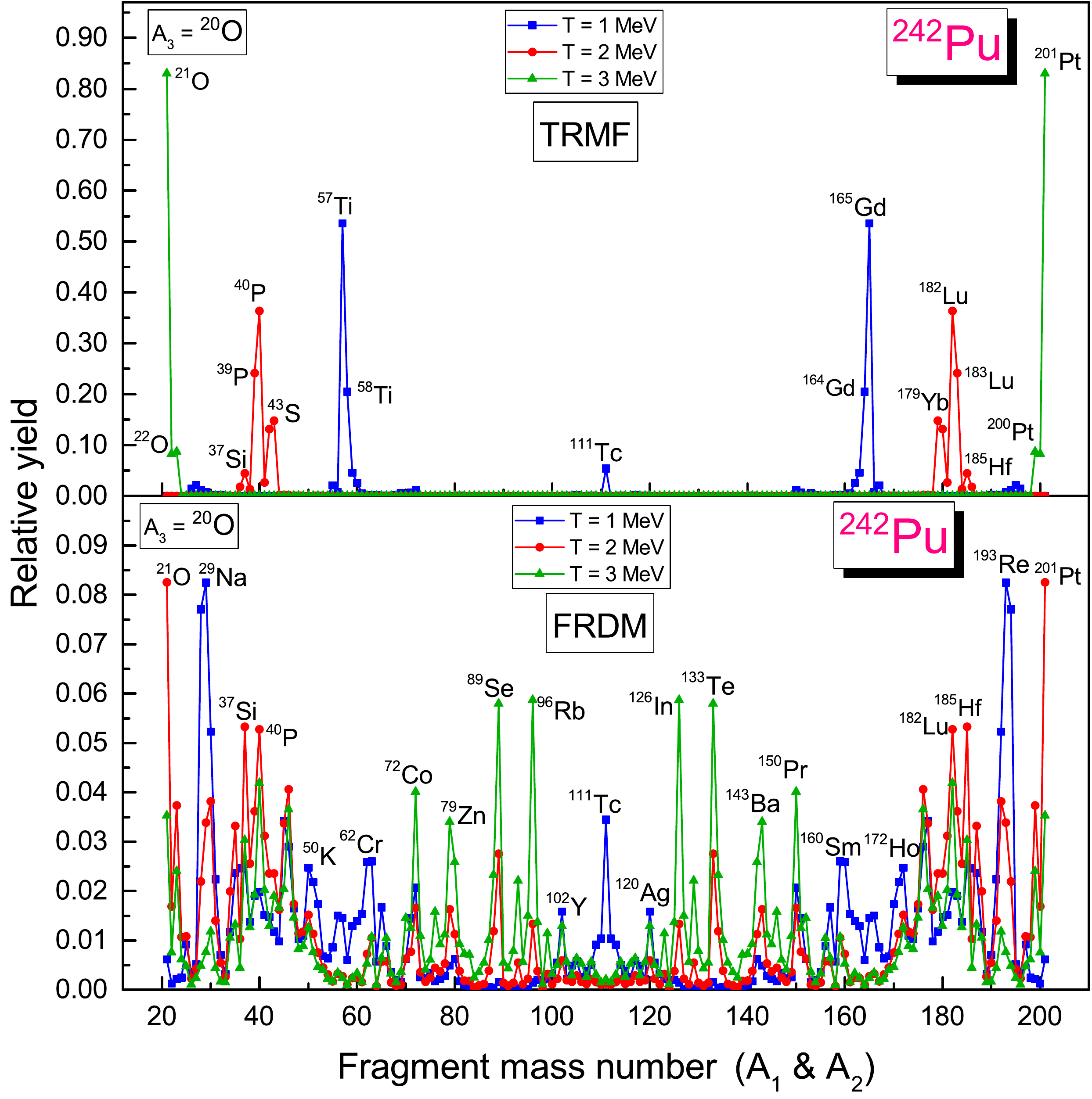}
\caption{(Color online) Mass distribution of $^{242}$Pu for the fixed
third fragment $^{20}$O for the temperatures T = 1, 2 and 3 MeV. The total yield values are normalized to 2.}
  	\label{Pu_yield}
  \end{figure}
  \begin{figure}[h]
  	\includegraphics[width=1.\columnwidth]{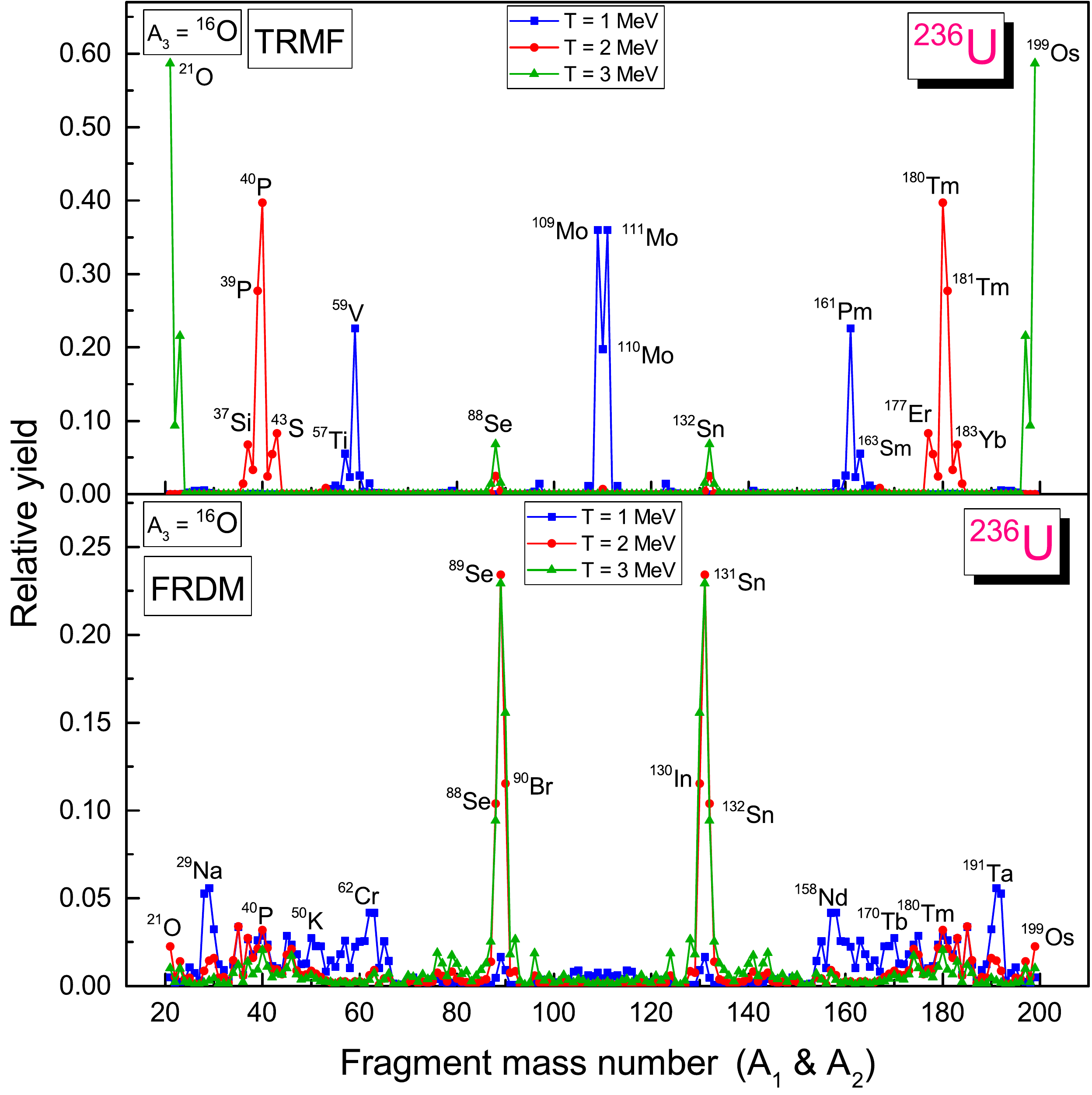}
  	\caption{(Color online) Mass distribution of $^{236}$U for the fixed third fragment $^{16}$O for the temperatures $T =$ 1, 2 and 3 MeV. The total yield values are normalized to 2.}
  	\label{U_yield}
  \end{figure}
In Figs.~\ref{Pu_yield} and \ref{U_yield}, we display the TRMF results for the ternary
fission mass distributions of $^{242}$Pu and $^{236}$U for the fixed
third fragments $^{20}$O and $^{16}$O respectively. At $T =$ 1 MeV, we see both symmetric and asymmetric yield for $^{242}$Pu and $^{236}$U. For the
$^{242}$Pu, at $T =$ 1 MeV, $^{165}$Gd $+^{57}$Ti $+^{20}$O is the most
favorable combination than the symmetric binary fragments $^{111}$Tc
$+^{111}$Tc $+^{20}$O. For $T =$ 2 MeV, $^{182,183}$Lu $+^{40,39}$P
$+^{20}$O, $^{179}$Yb $+^{43}$S $+^{20}$O and $^{185}$Hf $+^{37}$Si
$+^{20}$O are the possible relative yield. At $T =$ 3 MeV $^{201}$Pt
$+^{21}$O $+^{20}$O is the most favorable fragmentation. For $^{236}$U,
at $T =$ 1 MeV, the symmetric breakup into the heavy fragments $^{109}$Mo
$+^{111}$Mo and $^{110}$Mo $+^{110}$Mo along with the third fragment
$^{16}$O has larger yield values. In addition, the fragment combinations
$^{161}$Pm $+^{59}$V $+^{16}$O and $^{163}$Sm $+^{57}$Ti $+^{16}$O
also have larger yield values. For $T =$ 2 MeV, $^{180,181}$Tm
$+^{40,39}$P $+^{16}$O are the most probable fragments. Further,
$^{177}$Er $+^{43}$S $+^{16}$O and $^{183}$Yb $+^{37}$Si $+^{16}$O 
are also the probable ternary fragments.
It is seen that, at $T =$ 3 MeV, the fragments
$^{199,198,197}$Os $+^{21,22,23}$O $+^{16}$O have considerable yield
values. In Ref. \cite{sk16}, it is predicted that the ternary charge
distribution of $^{252}$Cf, at T$ =$ 2 MeV, with Si, P, S as the most
favorable fragments along with Sn and the corresponding partner. Here,
at $T =$ 3 MeV, the most favorable fragment is one of the closed shell ($Z = 8$) nucleus. Although, one would expect the even-even fragments as more probable for fission, we find large number of odd mass fragments possessing maximum yield compared to even-even. This is due to the fact that the level density of the odd mass fragments are higher than the even mass fragments as reported in Ref. \cite{fon56}.

\begin{figure}[t]
	\includegraphics[width=1\columnwidth]{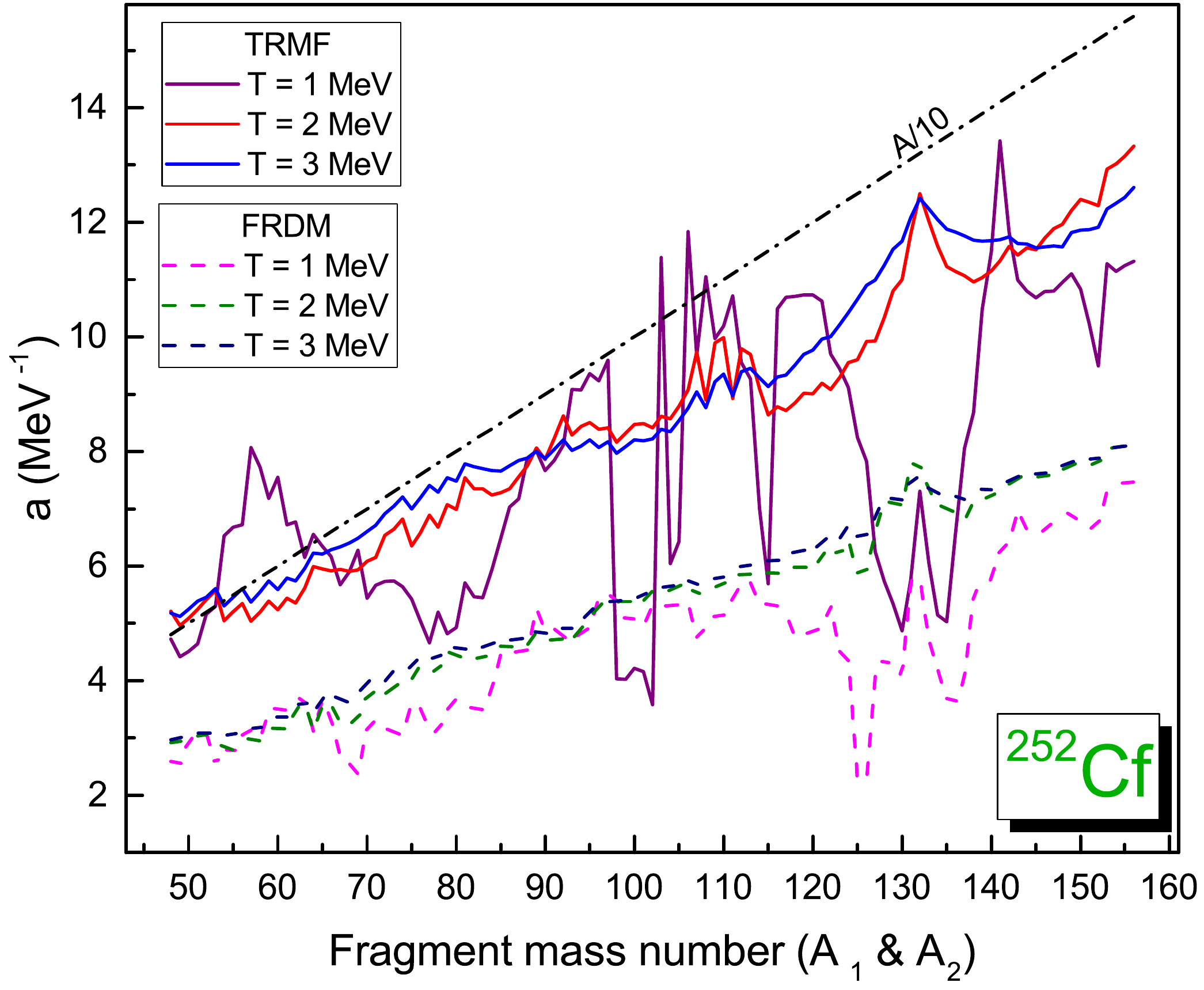}
	\caption{(Color online) The level density parameter $a$ of the ternary fragmentation of $^{252}$Cf for the temperature $T =$ 1, 2 and 3 MeV within the TRMF and FRDM formalism.
	}
	\label{252_LDP}
\end{figure}

 \begin{figure}[b]
 	\includegraphics[width=1\columnwidth]{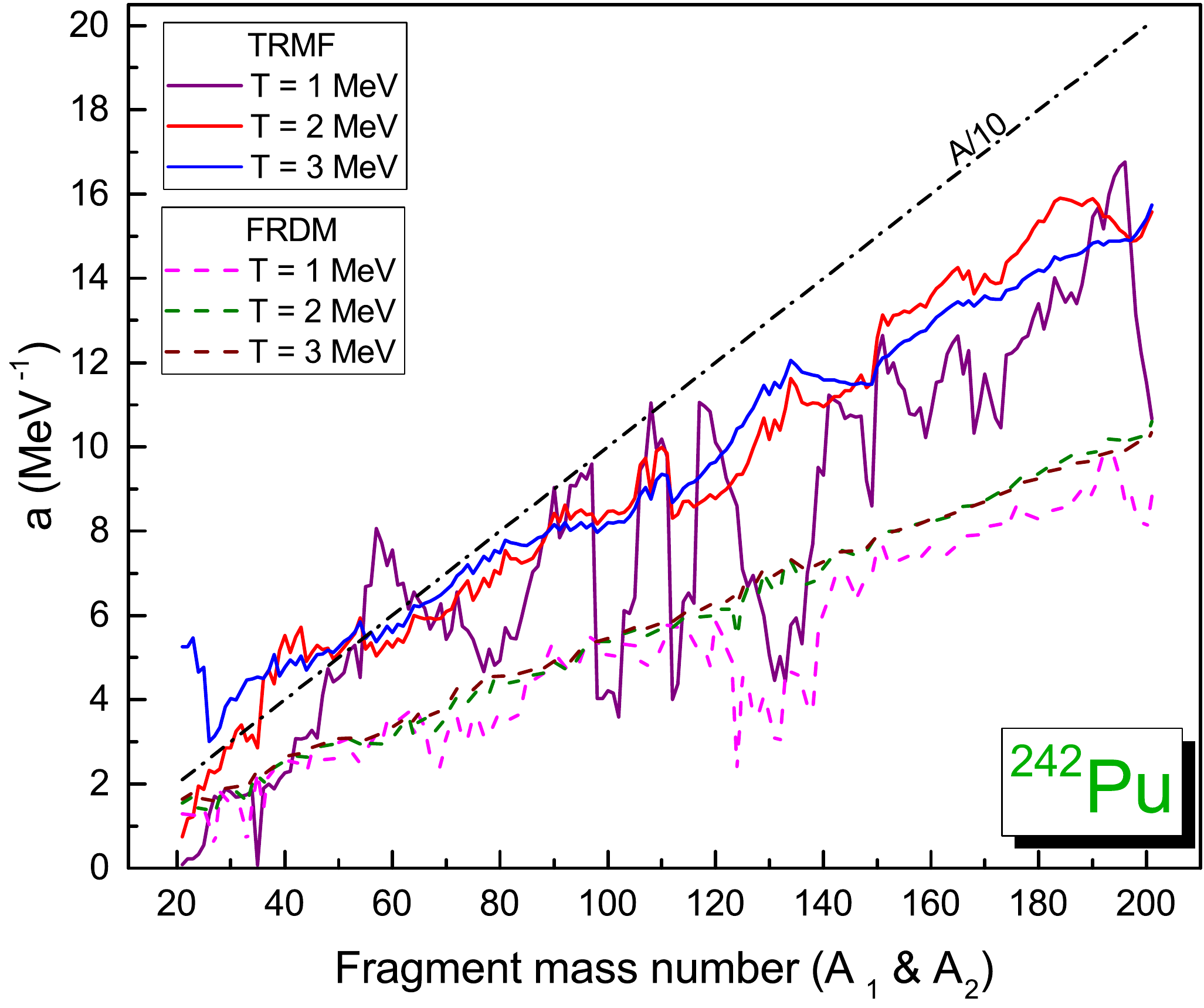}
 	\caption{(Color online) The level density parameter $a$ of the ternary fragmentation of $^{242}$Pu for the temperature $T =$ 1, 2 and 3 MeV within the TRMF and FRDM formalism.
 	}
 	\label{242_LDP}
 \end{figure}
 \begin{figure}[t]
 	\includegraphics[width=1\columnwidth]{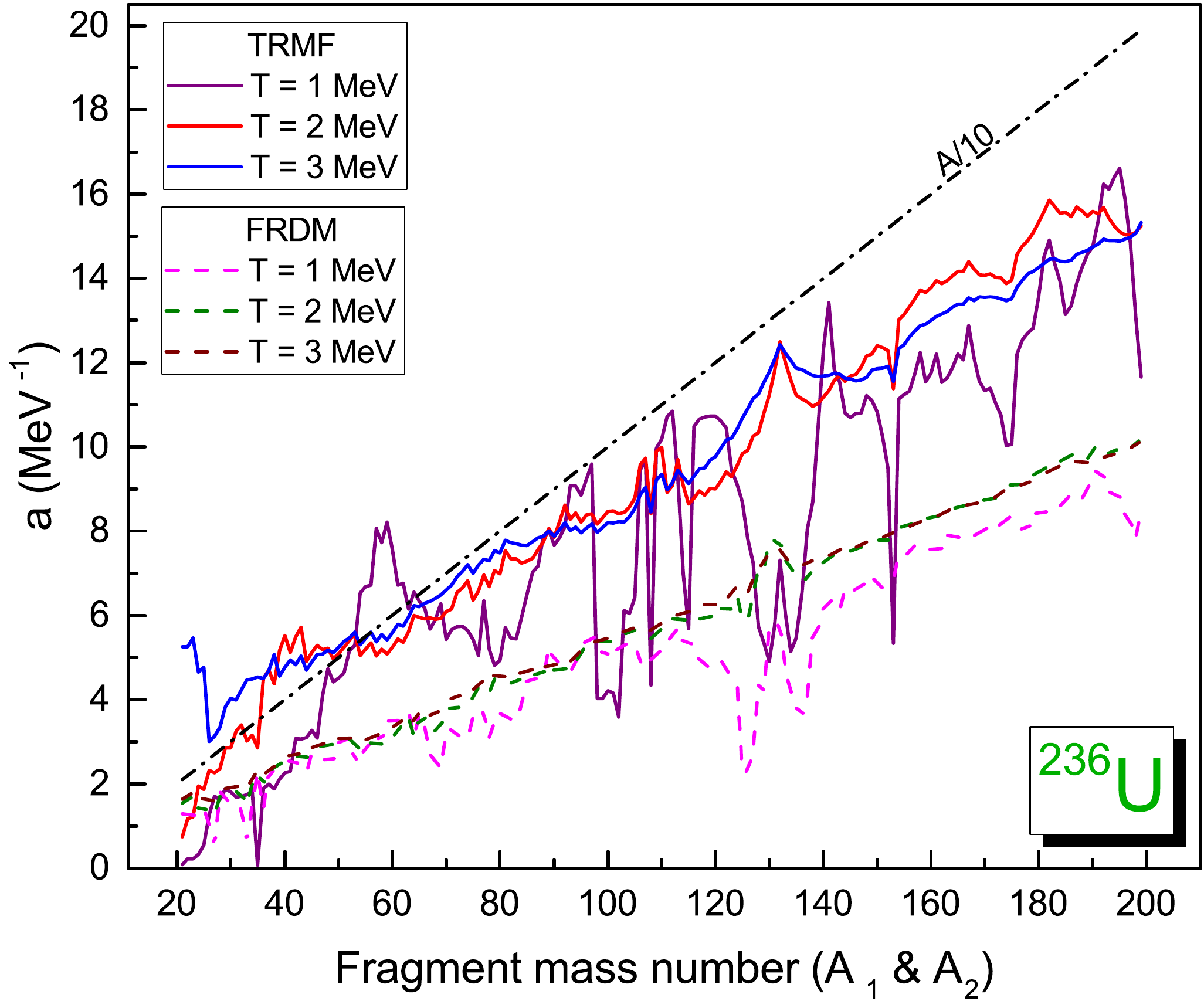}
 	\caption{(Color online) The level density parameter $a$ of the ternary fragmentation of $^{236}$U for the temperature $T =$ 1, 2 and 3 MeV within the TRMF and FRDM formalism.
 	}
 	\label{236_LDP}
 \end{figure}

\par
For the comparison, in Figs. \ref{Cf_yield} to \ref{U_yield}, the FRDM
results for the ternary mass distributions are also presented.
For the quick reference, the most probable ternary fragmentations
and their relative yield values are tabulated in Table \ref{table}
at three different temperatures $T =$ 1, 2 and 3 MeV. In general,  at $T =$ 1 MeV, the most favorable fragments of
the FRDM formalism are quite different than those for the TRMF.
These  differences may be attributed  to the differences in
the excitation energies obtained in the TRMF and FRDM formalisms.
For $^{252}$Cf the TRMF and FRDM results agree qualitatively with each other at $T =$ 2 and 3
MeV. For $^{242}$Pu more fragments have considerable yield values in
FRDM formalism. At $T =$ 2 and 3 MeV, the favorable fragmentations are
in the mass range $A_1 \sim$ 180 and 130 region with their corresponding
partners. The TRMF and FRDM results agree only partially
for the $^{242}$Pu nucleus at $T =$ 2 MeV. For $^{236}$U, the most
favorable fragments are at $A_1 \sim$ 130 for $T =$ 2 and 3 MeV in FRDM
calculations. One of the favorable fragments has a closed shell nucleon
or near closed shell ($N =$ 82) nucleus. Further, in
both the formalisms, at $T =$ 2 MeV, we get nearly similar yield such as $^{40}$P along with their partners $^{180}$Tm and $^{16}$O as shown in Fig. \ref{U_yield}. The doubly closed
shell nucleus $^{132}$Sn is appears in both the cases, at $T =$ 3
MeV.

\begin{figure}[b]
	\includegraphics[width=1\columnwidth]{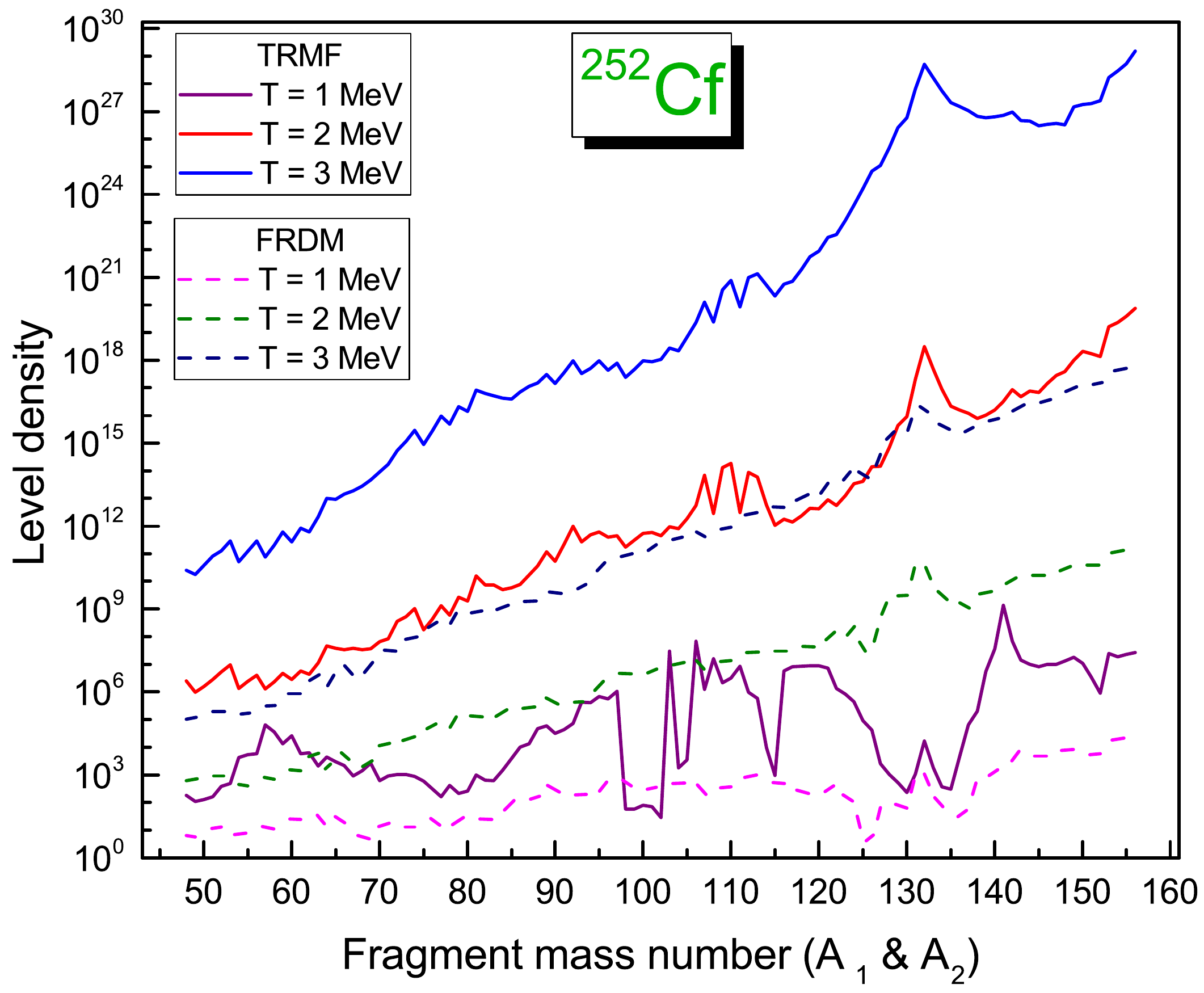}
	\caption{(Color online) The level density of the ternary fragmentation of $^{252}$Cf for the temperature $T =$ 1, 2 and 3 MeV.
	}
	\label{252_LD}
\end{figure}
\begin{figure}[h]
	\includegraphics[width=1\columnwidth]{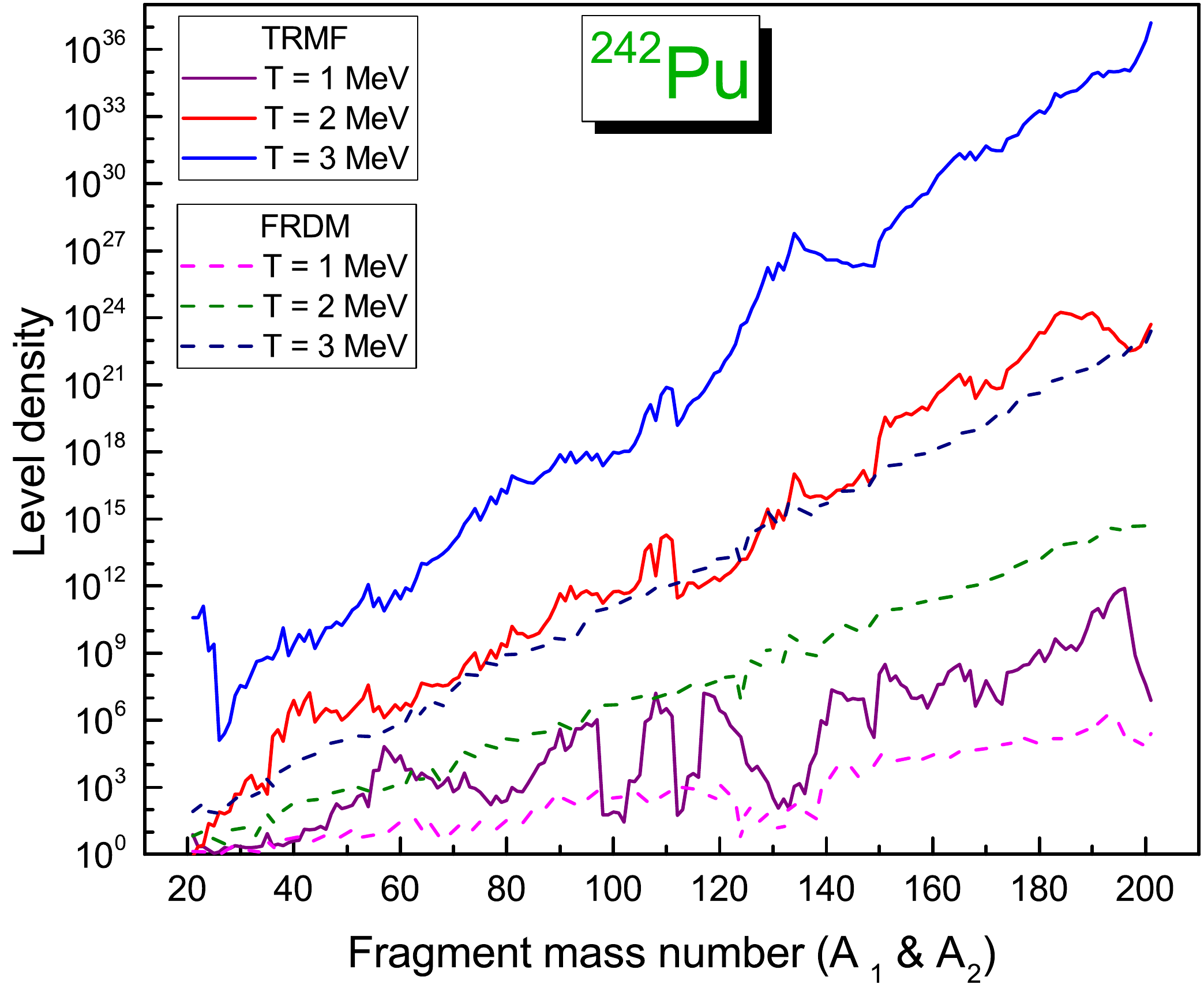}
	\caption{(Color online) The level density of the ternary fragmentation of $^{242}$Pu for the temperature T = 1, 2 and 3 MeV.
	}
	\label{242_LD}
\end{figure}

\par To illustrate the difference between the TRMF and FRDM results, we studied the level density parameter $a$ which is a crucial quantity. In general, the level density parameter $a$ is given by the empirical estimation relation \cite{ner2002}:
	\begin{equation}
		a = \dfrac{A}{K} (MeV^{-1}),
	\end{equation}
	
\begin{figure}[h]
	\includegraphics[width=1\columnwidth]{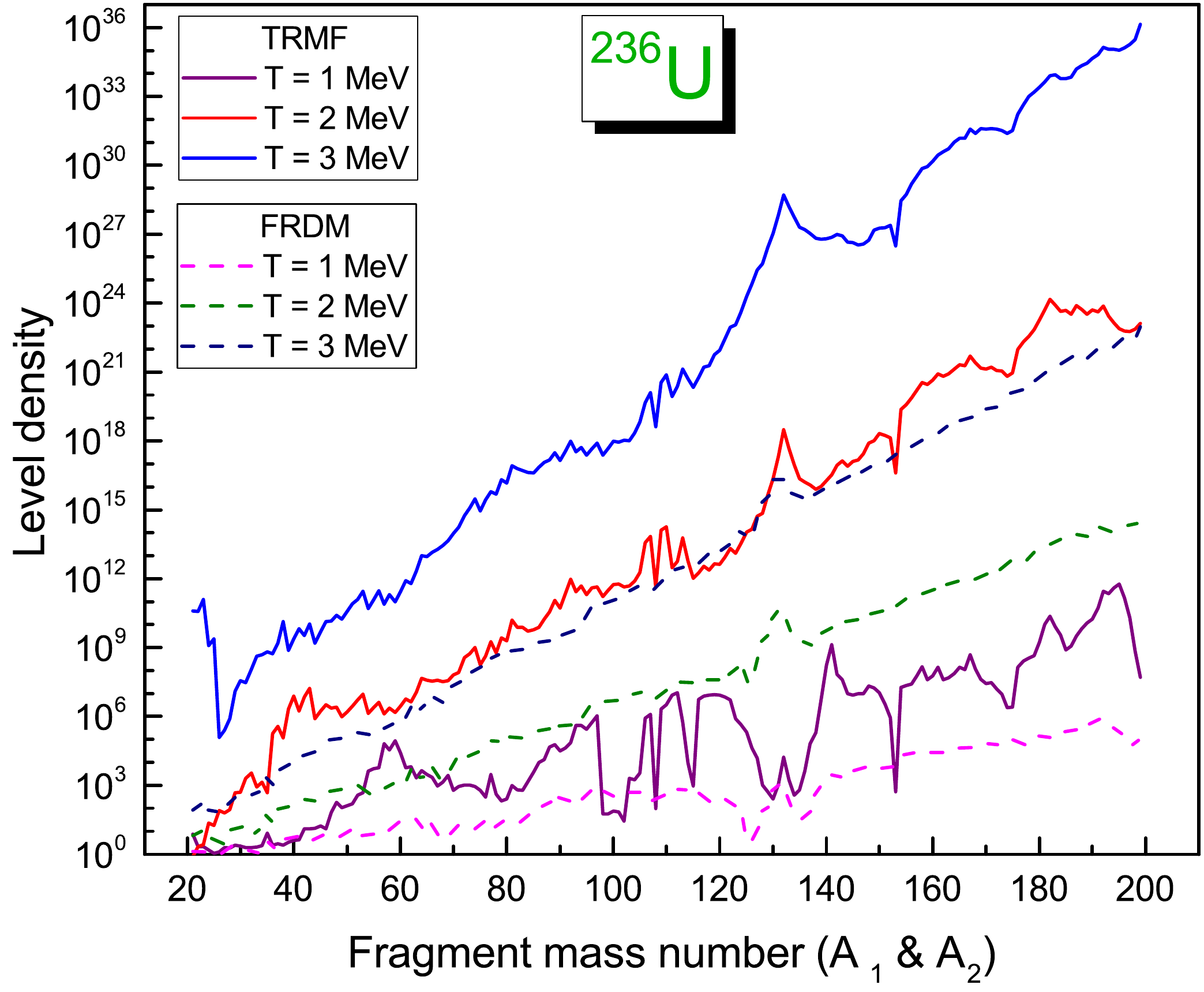}
	\caption{(Color online) The level density of the ternary
	fragmentation of $^{236}$U for the temperature $T =$ 1, 2 and
	3 MeV.	} \label{236_LD}
\end{figure}

\begin{table*}
	\caption{The relative fission yield (R.Y.) = $Y(A_j,Z_j)=\dfrac{P(A_j,Z_j)}{\sum P(A_j,Z_j)}$ for $^{252}$Cf, $^{242}$Pu and $^{236}$U obtained with TRMF at the temperatures $T= $1, 2 and 3 MeV are 
		compared with the FRDM prediction (The yield values are normalized to 2).} \label{table} 
	\centering
	\renewcommand{\arraystretch}{1.2}
	
	\begin{tabular}{|c|c|cc||cc|} 
		
		\hline
		
		\multirow{2}{1cm}{Parent} & \multirow{2}{*}{T (MeV)}  & \multicolumn{2}{c||}{TRMF} & \multicolumn{2}{c|}{FRDM }\\\cline{3-6}
		
		&&Fragment &R.Y.& Fragment & R.Y.\\
		\hline
		\multirow{6}{*}{$^{252}$Cf}&\multirow{2}{*}{1}& $^{108}$Nb + $^{96}$Rb + $^{48}$Ca &1.090&  $^{106}$Zr + $^{98}$Sr + $^{48}$Ca &0.196\\
		
		& 	& $^{141}$Xe + $^{63}$Cr + $^{48}$Ca &0.270& $^{113}$Tc + $^{91}$Br + $^{48}$Ca &0.134\\
		
		\cline{2-6}
		
		&\multirow{2}{*}{2}& $^{132}$Sn + $^{72}$Ni + $^{48}$Ca &1.438& $^{131}$Sn + $^{73}$Ni + $^{48}$Ca  &0.732\\	
		& 	& $^{160}$Nd + $^{48}$Ar + $^{48}$Ca &0.152& $^{132}$Sn + $^{72}$Ni + $^{48}$Ca &0.392\\	
		
		\cline{2-6}
		&\multirow{2}{*}{3}& $^{132}$Sn + $^{72}$Ni + $^{48}$Ca &1.508& $^{131}$Sn + $^{73}$Ni + $^{48}$Ca  &0.780\\	
		& & $^{131}$Sn + $^{73}$Ni + $^{48}$Ca &0.327& $^{128}$In + $^{76}$Cu + $^{48}$Ca &0.404\\\hline

		\hline 
		
		\multirow{9}{*}{$^{242}$Pu}&\multirow{3}{*}{1}& $^{165}$Gd + $^{57}$Ti + $^{20}$O &1.071&  $^{193}$Re + $^{29}$Na + $^{20}$O &0.165\\
		
		& 	& $^{164}$Gd + $^{58}$Ti + $^{20}$O &0.409& $^{111}$Tc + $^{111}$Tc + $^{20}$O &0.069\\
		&& $^{111}$Tc + $^{111}$Tc + $^{20}$O &0.107& $^{160}$Sm + $^{62}$Cr + $^{20}$O  &0.052\\	
		\cline{2-6}
		
		&\multirow{4}{*}{2}& $^{182}$Lu + $^{40}$P + $^{20}$O &0.726& $^{201}$Pt + $^{21}$O + $^{20}$O  &0.164\\	
		& 	& $^{183}$Lu + $^{39}$P + $^{20}$O &0.482& $^{185}$Hf + $^{37}$Si + $^{20}$O &0.106\\
		
		&& $^{179}$Yb + $^{43}$S + $^{20}$O &0.296& $^{182}$Lu + $^{40}$P + $^{20}$O  &0.106\\
		
		&& $^{185}$Hf + $^{37}$Si + $^{20}$O &0.090& $^{113}$Te + $^{89}$Se + $^{20}$O  &0.056\\	
		\cline{2-6}

		&\multirow{2}{*}{3}& $^{201}$Pt + $^{21}$O + $^{20}$O &1.660& $^{126}$In + $^{96}$Rb + $^{20}$O  &0.118\\	
		& 	& $^{200}$Pt + $^{22}$O + $^{20}$O &0.166& $^{133}$Te + $^{89}$Se + $^{20}$O &0.116\\\hline

		\hline 

		\multirow{10}{*}{$^{236}$U}&\multirow{3}{*}{1}& $^{111}$Mo + $^{109}$Mo + $^{16}$O &0.720&  $^{191}$Ta + $^{29}$Na + $^{16}$O &0.112\\
		
		& 	& $^{161}$Pm + $^{59}$V + $^{16}$O &0.452& $^{158}$Nd + $^{62}$Cr + $^{16}$O &0.084\\
		
		&& $^{110}$Mo + $^{110}$Mo + $^{16}$O &0.395& $^{180}$Tm + $^{40}$P + $^{16}$O  &0.058\\\cline{2-6}

		&\multirow{4}{*}{2}& $^{180}$Tm + $^{40}$P + $^{16}$O &0.794& $^{131}$Sn + $^{89}$Se + $^{16}$O  &0.468\\	
		
		& &	$^{181}$Tm + $^{39}$P + $^{16}$O &0.554& $^{130}$In + $^{90}$Br + $^{16}$O &0.230\\
		
		&& $^{177}$Er + $^{43}$S + $^{16}$O &0.166& $^{180}$Tm + $^{40}$P + $^{16}$O  &0.064\\
		
		&& $^{183}$Yb + $^{37}$Si + $^{16}$O &0.136& $^{199}$Os + $^{21}$O + $^{16}$O  &0.046\\	\cline{2-6}
		
		&\multirow{3}{*}{3}& $^{199}$Os + $^{21}$O + $^{16}$O &1.174& $^{131}$Sn + $^{89}$Se + $^{16}$O  &0.458\\	
		
		& 	& $^{197}$Os + $^{23}$O + $^{16}$O &0.432& $^{130}$Sn + $^{90}$Br + $^{16}$O &0.312\\	
		
		&& $^{132}$Sn + $^{88}$Se + $^{16}$O &0.136& $^{132}$Sn + $^{88}$Se + $^{16}$O  &0.188\\\hline

		\hline
		
	\end{tabular}
	
\end{table*}

where $K$ is the inverse level density parameter, varies from 10 to 14
depending on the mass number $A$ of the nucleus. In Figs. \ref{252_LDP}
- \ref{236_LDP}, we have plotted the level density parameter $a$
of the fission fragments for $^{252}$Cf, $^{242}$Pu and $^{236}$U
as a function of mass number. Here, we consider the inverse
level density parameter $K = 10$ (which is quite practical value
as mentioned in Ref. \cite{ner2002}) for all nuclei and shown
in the plots as black dashed dotted line. From these figures,
one can see that the TRMF values are very near to the empirical
level density parameter $a$. The FRDM values are considerably
lower than the referenced level density parameter. Further, in
both models at $T =$ 1 MeV, there are more fluctuations in $a$
due to the shell effects of the fission fragments. For $^{252}$Cf
and $^{236}$U, the level density parameter $a$ promptly increases
for the doubly closed shell nucleus $^{132}$Sn and has the lowest
inverse level density parameter $K = 10.9$. For $^{242}$Pu, the
$^{132}$Sn nucleus was restricted by Eq. \eqref{eq1}. However, value of
parameter $a$  increases towards the neutron closed shell ($N = 82$)
nuclei. In TRMF model the prompt increase of level density towards the doubly closed shell nucleus $^{132}$Sn are
clearly seen at $T =$ 3 MeV due to the fact all fission fragments becomes
spherical Fermi liquid drop as shown in Fig. \ref{spe}. 

\par To understand the results better we have plotted the level
	density of the fragments ($A_2$ and $A_1$) of the heavy nuclei $^{252}$Cf, $^{242}$Pu and
	$^{236}$U as a function of mass number as shown in Figs. \ref{252_LD}
	- \ref{236_LD}. From Fig. \ref{252_LD}, it can be seen that for $T =
	$ 2 and 3 MeV, the level density of $^{132}$Sn is higher than those for
	the neighboring nuclei in both formalisms. Hence, $^{132}$Sn becomes the most favorable
	fragment.  Fig. \ref{236_LD} shows, once again that $^{132}$Sn has higher level
	density than those for the  neighboring nuclei, however, the corresponding partner
	has lower/nearly same level density with the neighboring nuclei in the TRMF model. For
	the nucleus $^{242}$Pu, $^{132}$Sn is restricted due to the charge
	to mass ratio. From Figs. \ref{242_LD} and \ref{236_LD}, we see that
	the fragments Si to S have large level density compared with the
	neighboring nucleus and the corresponding partners also have the similar
	behavior. At $T =$ 3 MeV, the light charged particles, $Z_2 =$ 8 has
	larger level density than the neighboring nuclei and its corresponding
	partners also have the similar behavior. In FRDM formalism, the level density of doubly closed shell nuclei $^{132}$Sn has larger value than the neighboring nuclei for $^{252}$Cf and $^{236}$U at $T =$ 2 and 3 MeV. For $^{242}$Pu, there is no prompt increase in level density due to the restricted fragment $^{132}$Sn by Eq. \eqref{eq1}.
	
	Further, from Figs. \ref{252_LD}
	and \ref{236_LD}, it can be seen that the level density promptly increases while reaching the
	doubly closed shell nucleus $^{132}$Sn in both formalisms. It is noted that, other than the
	light charged particles, $^{132}$Sn has the larger level density. This
	ascertains the fact that with larger the phase space the ternary
	combinations becomes more probable than the other ternary fragments.

\section{Summary and conclusions} \label{se4}

We have studied the mass distribution of ternary fission fragments in
$^{252}$Cf, $^{242}$Pu and $^{236}$U  nuclei within the statistical
theory.  Various inputs to the statistical theory,  like, the excitation
energies and the level density parameters for the different fission
fragments at a given temperature are calculated from  the TRMF model.
The ternary combinations for these nuclei are obtained from the charge
to mass ratio of the parent nuclei. For the comparison, the results
obtained using the FRDM inputs to the statistical theory are also
presented.
	
For the nucleus $^{252}$Cf we obtained the Sn + Ni + Ca as the most
probable ternary combination at the temperatures $T =$ 2 and 3 MeV.  For the
nuclei $^{242}$Pu and $^{236}$U, however, we obtained few different
fragmentations at $T =$ 2 and 3 MeV. For these nuclei,  at $T =$ 2 MeV, the
Si/P/S are the possible ternary fragments along with the corresponding
fragments.  For $T =$ 3 MeV, the oxygen  isotopes have the larger yield
values. The TRMF results for the $^{252}$Cf at $T = 2$ and 3 MeV
resemble
very well with those for the FRDM. Whereas, they strikingly differ
from each other at $T = 1$ MeV. In the case of $^{236}$U, the mass
distributions for the ternary fission fragments corresponding to the
TRMF and FRDM resemble each other only at $T=3$ MeV. For the
$^{242}$Pu nuclei, the mass distributions for the TRMF
model and FRDM are by and large at variance at all the temperatures
considered. Thus, it seems that the  mass distribution of the fission
fragments are quite sensitive to  the effects on the excitation energy
and the level density parameter due to  the thermal evolution of the
deformation and the single-particle energies. This aspects are treated
self-consistently within the TRMF model, while, ignored within the
later approach.

\section{Acknowledgment}
The author MTS acknowledge that the financial support from UGC-BSR
research grant award letter no. F.25-1/2014-15(BSR)7-307/2010/(BSR)
dated 05.11.2015 and IOP, Bhubhaneswar for the warm hospitality and for
providing the necessary computer facilities.

\end{document}